\documentstyle[12pt]{article}
\def\singlespace {\smallskipamount=3.75pt plus1pt minus1pt
                  \medskipamount=7.5pt plus2pt minus2pt
                  \bigskipamount=15pt plus4pt minus4pt
                  \normalbaselineskip=15pt plus0pt minus0pt
                  \normallineskip=1pt
                  \normallineskiplimit=0pt
                  \jot=3.75pt
                  {\def\smallskip {\vskip\smallskipamount}}
                  {\def\medskip   {\vskip\medskipamount}}
                  {\def\bigskip   {\vskip\bigskipamount}}
                  {\setbox\strutbox=\hbox{\vrule
                    height10.5pt depth4.5pt width 0pt}}
                  \parskip 7.5pt
                  \normalbaselines}
\def\middlespace {\smallskipamount=5.625pt plus1.5pt minus1.5pt
                  \medskipamount=11.25pt plus3pt minus3pt
                  \bigskipamount=22.5pt plus6pt minus6pt
                  \normalbaselineskip=22.5pt plus0pt minus0pt
                  \normallineskip=1pt
                  \normallineskiplimit=0pt
                  \jot=5.625pt
                  {\def\smallskip {\vskip\smallskipamount}}
                  {\def\medskip   {\vskip\medskipamount}}
                  {\def\bigskip   {\vskip\bigskipamount}}
                  {\setbox\strutbox=\hbox{\vrule
                    height15.75pt depth6.75pt width 0pt}}
                  \parskip 11.25pt
                  \normalbaselines}
\def\doublespace {\smallskipamount=7.5pt plus2pt minus2pt
                  \medskipamount=15pt plus4pt minus4pt
                  \bigskipamount=30pt plus8pt minus8pt
                  \normalbaselineskip=30pt plus0pt minus0pt
                  \normallineskip=2pt
                  \normallineskiplimit=0pt
                  \jot=7.5pt
                  {\def\smallskip {\vskip\smallskipamount}}
                  {\def\medskip   {\vskip\medskipamount}}
                  {\def\bigskip   {\vskip\bigskipamount}}
                  {\setbox\strutbox=\hbox{\vrule
                    height21.0pt depth9.0pt width 0pt}}
                  \parskip 15.0pt
                  \normalbaselines}

\begin{document}
\newcount\sectionnumber
\sectionnumber=0

\title{Neutrinoless Double Beta Decay and CP Violation}
\author{Patrick J. O'Donnell \\
Physics Department,\\
University of Toronto,\\
Toronto, Ontario M5S 1A7, Canada.\\
\\
and \\
\\
Utpal Sarkar\\
Theory Group,\\
Physical Research Laboratory,\\
Ahmedabad - 380009, India.}

\date{May 27, 1993}

\maketitle

\begin{abstract}

\singlespace

We study the relation between the Majorana neutrino
mass matrices and the neutrinoless double beta decay when CP is not
conserved. We give an explicit form of the decay rate in terms
of a rephasing invariant quantity and demonstrate that in the
presence of CP violation it is impossible to have vanishing neutrinoless
double beta decay in the case of two neutrino generations
(or when the third generation leptons do not mix with other
leptons and hence decouple).

\vskip .5in
\begin{flushleft}
{\bf UTPT-93-12}\\
\end{flushleft}

\end{abstract}

\thispagestyle{empty}
\newpage
\middlespace

In the literature the relation between the Majorana neutrino
mass matrices and the neutrinoless double beta decay has
been studied extensively \cite{review}. If the electron neutrino
is a Majorana particle or a pseudo-Dirac particle ({\it i.e.,}
the symmetry of the mass matrix is not the symmetry of the
weak interaction), then they can contribute to neutrinoless
double beta decay, while if it is a Dirac particle then there
is no neutrinoless double beta decay. A Dirac particle can be
viewed as a combination of two Majorana particles with equal
mass and opposite CP properties and their contribution to the
neutrinoless double beta decay cancel \cite{wolf}.
In this article we
shall study this relation between the Majorana mass matrix and
the neutrinoless double beta decay when CP is not conserved ({\it
i.e.,} the mass eigenstates are no longer CP eigenstates).

We shall first discuss the question of CP violation
\cite{cpviol,pal} in the  leptonic sector and then
study neutrinoless double beta decay in
theories with CP violation.  We consider $n$ left-handed
Majorana neutrinos ($\nu_{iL}$, i = 1, ..., n) and $n$ charged
leptons ($l_\alpha$, $\alpha$ = 1, ..., n). For any CP violation
to take place we require $n \geq 2$. For $n = 2$ there is
exactly one CP violating phase; the number of phases
increases with increasing $n$.  The charged current interaction
and the mass terms of the lagrangian are
\cite{pal},
\begin{equation}
{\cal L} = \frac{g}{\sqrt{2}} \overline{{l'}_{\alpha L}} \gamma^\mu \nu'_{iL}
W_\mu^- - \overline{{l'}_{\alpha L}} M'_{l \alpha \beta} l'_{\beta R} -
\overline{{\nu'^c}_{iR}} M'_{\nu ij} \nu'_{jL} + h.c.
\end{equation}

We can now diagonalize the mass matrices and in the diagonal
basis (denoted by letters without any prime) we have,
\begin{equation}
{\cal L} = \frac{g}{\sqrt{2}} \overline{{l}_{\alpha L}} \gamma^\mu
\nu_{i L} V_{\alpha i}
W_\mu^- - \overline{{l}_{\alpha L}} M_{l \alpha} \delta_{\alpha \beta}
l_{\beta R} -
\overline{{\nu^c}_{i R}} K^2_{i} M_{\nu i} \delta_{i j} \nu_{j L} + h.c.
\end{equation}
The bi-unitary transformation, \hbox{$E^\dagger_{\alpha \beta R}
M'_{l \beta \eta} E_{\eta \rho L} = M_{l \alpha} \delta_{\alpha
\rho}$}, diagonalises the charged lepton mass matrix,
 while
the symmetric neutrino mass matrix is diagonalized as,
\begin{equation}
U^T_{ij} M'_{\nu j k} U_{k l} = K^2_i M_{\nu i} \delta_{i l}
\end{equation}
where $M_\nu$ has real positive mass eigenvalues and $K$ is a
diagonal phase matrix. The mixing matrix $V$, given by
$V = E_L^\dagger U $, is analogous to the Kobayashi-Maskawa
matrix of the quark sector.

Under CP the Weyl fields transform as, $\nu_L
\to {\nu^c}_R = {\nu_L}^c$ and $\nu_R \to {\nu^c}_L = {\nu_R}^c$.
CPT invariance implies that the CP conjugate fields are always present
in the theory. Thus considering any Weyl field $\nu_L$ is equivalent
to reducing the number of degrees of freedom by half and in the
above the mass eigenstate Majorana fields $\nu_i = U^\dagger_{i j}
\nu'_{j L} + K^{*2}_{i} U^T_{ij} {\nu'^c}_{jR}$ satisfy the
Majorana condition $\nu = K^{*2} \nu^c$ so as to match the number of
degrees of freedom.
Under CP transformations all terms in the lagrangian go to
their hermitian conjugates except for the coefficients. As a result
if there is any complex coefficient, then the imaginary part gives
rise to CP violation. However, many of the phases can be rotated
away by the rephasing of the fermion fields, since physical
processes are independent of the rephasing of these fields.

In principle, both the matrices $V$ and $K$ can have CP
violating phases and can have observable  consequences
\cite{cpviol,pal}.
Under the rephasing of the neutrino fields, $\nu_i \to e^{i
\delta_i} \nu_i$, the invariance of the charged current interactions
and the Majorana mass terms dictates that $V$ and $K$ should
transform as
$$\{V_{\alpha i}, K_i\} \to e^{- i \delta_i} \{ V_{\alpha i},
K_i\}.$$
In general the $n-1$ phases of $K$ (the overall phase is
not significant) can be absorbed in the rephasing
of the $n$ fields $\nu_{i L}$. This will introduce
new phases in $V$. If there is no CP violation in the lagrangian
then these phases can be put to zero. With CP violation in the
lagrangian this rephasing will leave nonvanishing phases in
$V$. This is in contrast to what happens in
the quark sector, where the absence of a Majorana mass term
implies the phase rotation of the right handed fields can be
exploited to make $K^2$ real.

Although $V$ and $K$ both transform under the rephasing of
the fermion fields and are not invariant, in physical
processes they enter in certain combinations which are
rephasing invariant. Some of these invariant forms has
been studied in the literature \cite{pal}.
In neutrinoless double beta decay processes
only one of the rephasing invariant combinations enters
\begin{equation}
s_{\alpha i j} = V_{\alpha i} V^*_{\alpha j} K^*_i K_j,
\end{equation}
which has the following properties:
\begin{equation}
s_{\alpha i j} s_{\alpha j k} = |V_{\alpha j}|^2 s_{\alpha i k};
\;\;\:\:\: s_{\alpha i j} = s_{\alpha j i}^*  .
\end{equation}

If any of the $s_{\alpha i j}$ are complex
then that will imply CP violation in processes in which they enter. For
a given flavour $\alpha$, $s_{\alpha }$ is hermitian. Thus for
$n$ generations  $s_{\alpha }$ can have $n(n-1)/2$ phases.
For two neutrino generations (for  $\alpha = e \;\;{\rm and} \;\;\mu$)
there is one independent phase in  $s_{\alpha 1 2}$.

The amplitude for the neutrinoless  double beta decay arises
from the term,
\begin{equation}
\sum_i [W_\mu \overline{{l}_{\alpha L}} \gamma^\mu \nu_{i L}
V^*_{\alpha i}] \:\:\: [K_i^2
W_\nu \overline{{\nu}_{i L}} \gamma^\nu e^c_{\alpha L} V^*_{\alpha i}]
\end{equation}
with $\alpha = e$ and the decay width  is proportional to,
\begin{equation}
\Gamma \propto |\sum_i V^*_{e i} V^*_{e i} K_i^2 M_{\nu
i}|^2 = \sum_{i,j} s_{e i j}^2 M_{\nu  i} M_{\nu j}.
\end{equation}
The advantage of writing the decay rate in the rephasing invariant
form is that now we do not have to worry about the CP properties
of the neutrinos or the phases and the sign convention. Any result
given in this approach will automatically take care of the phase
convention. Thus, if there is CP
violation in the theory then some of the $s_{e i j}$ can be
complex and the decay rate will depend on the amount of CP
violation in the theory. This means that the CP violating phase
may not allow cancellation of different terms in the amplitude
\cite{wolf}. In other words, if CP is
conserved then two Majorana neutrinos of equal mass and opposite
CP eigenvalue can combine to form a Dirac neutrino \cite{maj};
however, if there is CP violation then the neutrinos no longer
form CP eigenstates.

We consider a two generation example to demonstrate how
CP violation changes the decay rate. For the charged lepton mass
matrix we take the mixing matrix to be of the form,
$$ E_L = \pmatrix{ c_e & -s_e \cr s_e & c_e}, $$
where, $c_\alpha = \cos \theta_\alpha$ and $s_\alpha = \sin \theta_\alpha$.
We shall consider the neutrino mass matrix to be the most general one.
Instead of considering any particular form for the mass matrix
$M'_\nu$, we start with the diagonal form of the mass matrix to
be  $M_\nu = \pmatrix{m_1  & 0 \cr 0 & m_2 } $, the mixing
matrix to be of the general form,
$U = \pmatrix{c_\nu & s_\nu \cr {-s_\nu} & c_\nu}$ and
$K = \pmatrix{1 & 0 \cr 0 & e^{i \delta} }$.
This will correspond to the most general form for
the mass matrix $M'_\nu$ except for some possible phase rotations. But
since we are dealing with the phase invariant quantities any
other phase choice will not change our result. We then get,
\begin{equation}
\Gamma \propto A^2 + B^2 + 2 A B \cos 2 \delta ,
\end{equation}
where, $A = (c_e c_\nu - s_e s_\nu)^2 m_1 $ and $B= (c_e s_\nu +
s_e c_\nu)^2 m_2 $.
Obviously the decay rate depends explicitly on the CP violating
phase $\delta$. This decay rate vanishes only for $A = B$ and
$\delta = \pi/2$, which corresponds to there being no CP violation. This
also implies that the two mass eigenstates have opposite CP
properties and their contributions to the decay rate cancels
each other \cite{wolf}. If there is CP violation then this cancellation
is incomplete and there is no other solution for which the decay
rate vanishes. Thus for two generations in the presence of CP
violation it is impossible to have vanishing neutrinoless double
beta decay.

However, if we work in the basis where the charged lepton mass matrix is
diagonal, {\it i.e.,} the weak eigenstate basis, (to be precise,
we require only $E_L = \pmatrix{1&0 \cr 0&1}$) then equation (7)
reduces to,
\begin{equation}
\Gamma \propto |M'_{\nu 1 1}|^2.
\end{equation}
Although this apparently contradicts our previous remarks, in
practice it does  not. When $M'_{\nu 1 1}$ or any one of the
three elements of $M'_\nu$ vanishes,
there is no CP violation.

To see this consider  the basis where $E_L$ is diagonal;
$E_L$  can then  be made real as well.
All the CP violating phases in the lagrangian are then contained
in the neutrino mass matrix $M'_\nu$.
Since $M'_\nu$ is symmetric only $[n^2 - n(n-1)/2]$
independent phases can be present,
of which we can absorb
$n$ phases by rephasing the $n$ neutrino fields, leaving only
$n(n-1)/2$ independent phases.
In the case of two generations this
means that only one of the three elements of $M'_\nu$ can be complex
and if any one of them is zero we can always make $M'_\nu$
real.

This conclusion may not be valid for the case of three
generations, where the total number of phases in $M'_\nu$ is
three. Thus if $M'_{\nu 1 1} = 0$, implying vanishing of the
neutrinoless double beta decay, the matrix $M'_\nu$ can still have
two more complex phases and hence even in the presence of CP
violation the contribution to the neutrinoless double beta decay
can vanish. However, in the three generation case if the third
generation decouples from the other two generations, {\it i.e.,}
the third generation leptons do not mix with others both in the
neutrino as well as in the charged lepton sector, then our
discussion with the two neutrino generations is still valid and
again it will be impossible to have vanishing neutrinoless
double beta decay with CP violation.

To summarize, we have studied the relation between the neutrinoless
double beta decay rate and the neutrino mass matrix when CP is
not conserved. In the two family scenario (or where the third
family decouples from the other two), it is not possible to have
non-zero neutrinoless double beta decay.

{\bf Acknowledgement} We would like to thank the
NSERC of Canada for an International Scientific Exchange
Award.

\end{document}